\documentclass{Interspeech}

\usepackage{graphicx}
\usepackage{amsmath}
\usepackage{soul}
\usepackage{adjustbox}

\interspeechcameraready

\title{Robust Target Speaker Diarization and Separation \\ via Augmented Speaker Embedding Sampling}

\author[affiliation={1}]{Md Asif}{Jalal}
\author[affiliation={1}]{Luca}{Remaggi}
\author[affiliation={2}]{Vasileios}{Moschopoulos}
\author[affiliation={2}]{Thanasis}{Kotsiopoulos}
\author[affiliation={1}]{Vandana}{Rajan}
\author[affiliation={1}]{Karthikeyan}{Saravanan}
\author[affiliation={2}]{Anastasis}{Drosou}
\author[affiliation={3}]{Junho}{Heo}
\author[affiliation={3}]{Hyuk}{Oh}
\author[affiliation={3}]{Seokyeong}{Jeong}

\affiliation{}{Samsung R\&D Institute UK (SRUK)}{United Kingdom}
\affiliation{}{Centre for Research and Technology Hellas}{Greece}
\affiliation{}{Language AI R\&D Group (MX), Samsung Electronics}{South Korea}
\email{mdasif.jalal@samsung.com}

\usepackage{comment}
\usepackage{lipsum}

\begin{document}

\maketitle


\begin{abstract}
\vspace{-1mm}
Traditional speech separation and speaker diarization approaches rely on prior knowledge of target speakers or a predetermined number of participants in audio signals. To address these limitations, recent advances focus on developing enrollment-free methods capable of identifying targets without explicit speaker labeling. This work introduces a new approach to train simultaneous speech separation and diarization using automatic identification of target speaker embeddings, within mixtures. Our proposed model employs a dual-stage training pipeline designed to learn robust speaker representation features that are resilient to background noise interference. Furthermore, we present an overlapping spectral loss function specifically tailored for enhancing diarization accuracy during overlapped speech frames. Experimental results show significant performance gains compared to the current SOTA baseline, achieving 71\% relative improvement in DER and 69\% in cpWER.
\end{abstract}

%
%
\vspace{-3mm}
\section{Introduction}
\vspace{-1mm}
\label{sec:intro}
In recent years, advancements in Automatic Speech Recognition (ASR)~\cite{KHEDDAR2024102422, Meng2024SEQformerAC, WangEfcientlyTA, YoonWooKin24}, have enabled pivotal use-cases such as language translation~\cite{ShenYuanShenDuYu2025}, human-computer interaction~\cite{Lyu01112024}, speech transcription~\cite{Chime8} and diarization~\cite{Kim2021NorthAB}. 
ASR performance degrades when dealing with recordings containing overlapping voices of multiple speakers. Such scenarios are prevalent in everyday situations like business meetings, casual conversations, and even at noisy gatherings, often referred to as the ``cocktail party'' effect \cite{CocktailParty}. A leading solution involves the introduction of a speech separation module prior to the ASR~\cite{9383615, 10.1109/TASLP.2019.2915167, 9746372, Shin2024SeparateAR}. This pre-processing step aims to separate speakers' voices into individual channels prior to the transcription. However, traditional speech separation methods come with inherent limitations: they require advance knowledge of the number of speakers within a recording; and they suffer from the ``permutation problem'', where speaker's identity is not consistently assigned to the same output channel~\cite{7952154}. 

Recent advances in Target Speaker Separation~\cite{wang2019voicefilter, SpeakerBeam19,  JiuxinPenHeinJunZhiZhiZhaYuj2023, chetupalli2024unified, Sato2024, He2021TargetSpeakerVA, boeddeker2024ts, Boedd2024-is} address these challenges. They use prior knowledge about the target speaker(s) to be separated, specifically, in the form of speaker embeddings, to describe the characteristics of the speaker voice(s)~\cite{10760244}. Typically, the embeddings are obtained by processing pre-recorded enrollments from target speaker(s) through a speaker encoder model~\cite{wang2019voicefilter, zeng2025universalspeakerembeddingfree}. While effective, relying on pre-enrollments imposes constraints on the versatility of the speech separation-ASR workflow, thus limiting its application to scenarios wherein these prerequisite recordings exist. He \textit{et al}. \cite{He2021TargetSpeakerVA} showed impressive diarization results with unknown number of speakers, while Boeddeker \textit{et al}. \cite{boeddeker2024ts} found that some compromise in terms of DER needs to be made to achieve great ASR results, simultaneously.


This paper draws inspiration of enrollment-free target speaker separation, from the two-stage TS-SEP framework introduced by Boeddeker \textit{et al}. \cite{boeddeker2024ts}. In TS-SEP, a diarization system determines the speakers in the audio recording without requiring prior speaker enrollment data, and extracts individual speaker embeddings for target speech separation. 
It was observed that the final speech separation results were sensitive to the quality of the estimated speaker embeddings \cite{boeddeker2024ts}. The difference in the the way speaker embeddings were obtained during training and evaluation could be a possible reason. During training, the speaker embeddings were obtained from clean speech with no overlap, and during evaluation, the chosen target speaker segments often contained overlapping superimposed speech segments (i.e., overlapping speakers). Therefore, the models needs to learn robust generalised understanding of target speaker among overlapping speaker mixtures. In this paper, we hypothesize that, pre-training the speech separation model towards noisy Target Speaker Voice Activity Detection (TS-VAD)~\cite{TSVAD} task (Stage 1), by sampling noisy (speaker overlapped) embeddings, would significantly improve speech separation and diarization tasks (Stage 2).
 This hypothesis is similar to controlled noisy pretraining, which degrades the performance of upstream task, to improve generalisation and performance in the downstream task \cite{51104, chen2024learning, pmlr-v97-hendrycks19a}. For example, training a vision transformer with noisy images and contrastive loss improves generalisation on downstream tasks \cite{chen2024learning}.

Our contributions are the following: 
\begin{enumerate}
    \item A novel sampling procedure identifying speech frames with highest variability in background noise, overlapped speakers.
    \item A speaker-overlapped noisy pre-training for speaker dependent Voice Activity Detection (VAD), which is used to initialise the speech separation model.
    \item A method to address the background speech (or overlapped speech), with a robust speaker encoder based clustering and improved Diarization Error Rate (DER) by relative 71.3\%.
    \item At inference, included an overlapped speech detection model acting as a mixture of experts along with the clustering for sourcing embedding from single speaker speech segments.
    \item Overlapping-window spectral reconstruction loss improves temporal coherence of the separated speech, by 22\% concatenated minimum-permutation Word Error Rate (cpWER) \cite{meeteval}.
\end{enumerate}

The rest of the paper is structured as follows:
Section~\ref{sec:method} discusses our proposed methods while framing it into the target speaker separation application. Section~\ref{sec:experiment} describes the objectives of our experiments and our setup. In Section~\ref{sec:results} we show the experimental results, discuss them and describe their entailment. Finally, Section~\ref{sec:conclusion} draws the conclusions of our findings.



\begin{figure*}[ht!]
\centering
\includegraphics[height=0.18\linewidth,width=\textwidth]{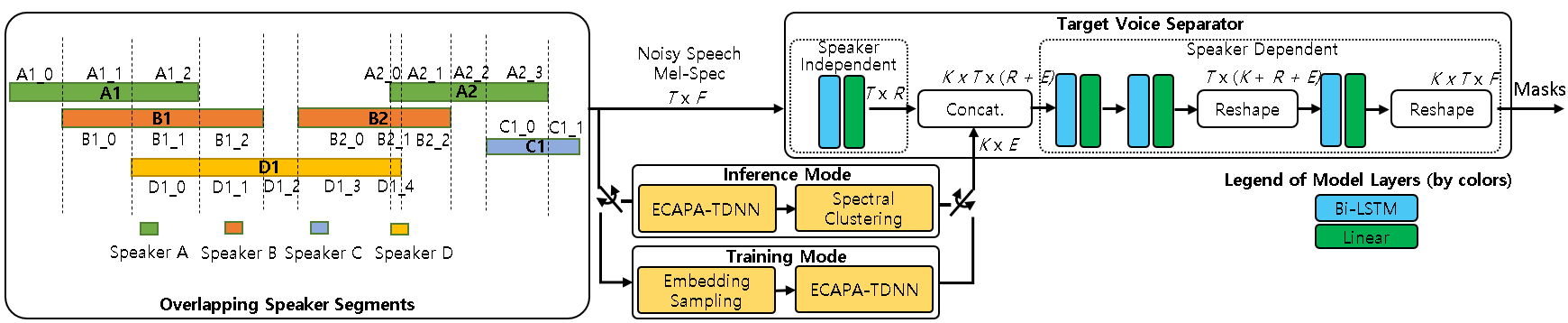}
\caption{Proposed Model Pipeline and Utterance Segmentation. In a mix of four speakers, defined as $U \in [A, B, C, D]$, some segments $U_g$ overlap with others, where $g$ is the segment index. Their sub-segments are denoted as $U_{g,b}$, where $b$ is the sub-segment index. $T$ is the number of time frames, $F$ is the number of frequency bins, $K$ is the number of speakers in the mixture, $E$ is the speaker embedding dimensions, and $R$ is the number of dimensions in the latent space.}
\label{fig:embed_aug}
\vspace{-3mm}
\end{figure*}

\section{Method}
\label{sec:method}
\vspace{-1mm}

\subsection{Multi-Stage Training}
\vspace{-2mm}


The training strategy is composed of \cite{boeddeker2024ts}: Stage 1, which trains a speaker-dependent VAD model~\cite{TSVAD}; Stage 2, which trains the target speech separation model starting from Stage 1. Let $X(f,t) \in \mathbb{R}^{T \times F}$ be the time-frequency Short Time Fourier Transform (STFT) representation of the input mixture, where $T$ and $F$ are the time frames and frequency bins, respectively, having indexes $t$ and $f$. Every speaker $k$ is associated with an embedding $I_{k} \in  \mathbb{R}^{E}$, where ${E}$ is the embedding dimension.
\vspace{-2mm}
\subsubsection{Stage 1 Training}
\label{sec:stage1}
\vspace{-2mm}

The speaker-dependent VAD model predicts a vector $V_{pred, k}(t) \in [0,1]^{T}$, indicating at each frame $t$ whether speaker $k$ is active or not. 
It is trained on the VAD task for each of the $K$ speakers in the mixture, by optimizing the binary cross entropy loss between the predictions $V_{pred, k}(t) \in [0,1]^{T}$ of the model and the ground truth labels $V_{gt, k}(t) \in [0,1]^{T}$. 

\vspace{-2mm}
\subsubsection{Stage 2 Training}
\label{sec:stage2}
\vspace{-2mm}
To initialise the speech separation model from the speaker-dependent VAD model at Stage 1, its final layer is replicated $F$ times, yielding time-frequency output masks, defined as $M_k(f,t) \in [0,1]^{T \times F}$. To obtain the separated time-frequency representation of the signal for each speaker $\hat{Y}_{k}(f,t) \in \mathbb{R}^{T \times F}$, each mask $M_{k}(f,t)$ is combined by the single channel mixed input via Hadamard product, formally ${\hat{Y}_{k}}(f,t) = M_{k}(f,t) \odot X(f,t)$.
The speech separation model is trained by employing the following time domain reconstruction loss:
\vspace{-3mm}
\begin{equation}
\mathcal{L}_{SEP} = log_{10}\biggr[\frac{1}{KN} \sum_{k=1}^{K} \sum_{n=1}^{N} |\hat{y}_k(n) - y_k(n)|\biggr]
\label{eq:recons}
\end{equation}
\vspace{-4mm}

where $| \cdot |$ is the absolute value operator, $y_{k}(n)$ is the groundtruth time domain signal and $\hat{y}_{k}(n)$ is the inverse STFT to ${\hat{Y}_{k}(f,t)}$. $n$ is the discrete time domain sample index and $N$ is the total number of time domain samples composing the signal.

\vspace{-2mm}
\subsection{Speaker Encoder}
\vspace{-2mm}
\label{sec:spk_emb}
Both speaker-dependent VAD and speech separation rely on high-quality speaker embeddings. In Boeddeker \textit{et al}. \cite{boeddeker2024ts}, x-vector was utilized to derive embeddings from speech segments. From them, Spectral Clustering~(SC) forms distinct speaker clusters. Only those clusters comprising single speakers were then selected and i-vectors were calculated from the corresponding speech excerpts,which in turn serve as the input to the separation model. Notably, Boeddeker \textit{et al}. \cite{boeddeker2024ts} asserted that i-vector outperformed x-vector in both VAD and speech separation tasks. However, it is known that for speaker verification, the SOTA ECAPA-TDNN architecture outperforms both i-vector and x-vector methods \cite{desplanques2020ecapatdnn}. Therefore herein, we employ a unified framework leveraging a compact six-million-parameter ECAPA-TDNN architecture~($e6m$) \cite{desplanques2020ecapatdnn}, exclusively dedicated to generate robust speaker representations (Figure \ref{fig:embed_aug}).

\vspace{-2mm}
\subsection{Noisy Embedding Augmentation and Sampling}
\label{sec:noisy_embed}
\vspace{-2mm}

Enrollment-free joint diarization and separation perfomance is affected by noisy and overlapped speaker embeddings~\cite{bullock:hal-02995367}. Since speech separation models are trained on target-speaker embeddings (without overlaps), their performance on overlapped speech is often poor, leading to incorrect speech separations~\cite{boeddeker2024ts}. This is evident from the difference in evaluation results between oracle and non-oracle embeddings (Table \ref{tab:ablation}). Therefore, we propose a method where a pre-trained model is trained on noisy embeddings for speaker-dependent voice activity detection. We hypothesise and empirically show that the acquired knowledge through this training scheme aids generalization of the speech separation model, which $s_\theta$ is defined as:
\vspace{-2mm}
\begin{equation}
    s_\theta\left(task_{s}, d_\theta\left( task_{v}, \mathcal{A}, I_{noisy} \right), I  \right)
\end{equation}

\vspace{-1mm}

where $d_\theta$ is the speaker-dependent VAD model, $\mathcal{A}$ is the speaker-dependent VAD model architecture, $task$ is the training task (s:separation; v:vad), $I$ is the speaker embedding, and $I_{noisy}$ is the noisy speaker embedding. We opted to train with speaker-overlapped noisy embeddings, with up to 10\% speaker overlap and silence. The degree of noise augmentation was carefully regulated since controlled noisy pre-training enhances generalized feature elements \cite{chen2024learning}. However, excessive noise can lead to the model overfitting to the noise patterns~\cite{chen2024learning, pmlr-v97-hendrycks19a}. 
 The left side of figure \ref{fig:embed_aug} depicts the overlapping speakers in an audio mixture, and emphasizes the overlap and non-overlap segments. The top right represents the TS-SEP model \cite{boeddeker2024ts}. The bottom centre shows the procedure to generate speaker embeddings for TS-SEP. During training, we utilize the proposed embedding sampling. If $SE$ is the speaker encoder, $\mu$ is the mean and \emph{silence} is silent frames, then the sampling methods are:

\textbf{V1}: V1 is the oracle embedding for each speaker in the given mixture. Example, $V1 = \mu(SE(A1)+SE(A2))$.

\textbf{V2}: V2 is the average of the non-overlapping speech segments for each speaker in the given mixture. Example, $V2 = \mu(SE(A1\_0) + SE(A2\_2))$.

\textbf{V3}: V3 is prepared by concatenating silence with the non-overlap segments. Example, $V3 = \mu(SE(A1\_0 + silence) + SE(silence+A2\_2+silence))$.

\textbf{V4}: V4 is V2 with added 10\% overlapped segment regions. Example, $V4 = \mu(SE(A1\_0+10\%(A1\_1+B1\_0)) + SE(A2\_2 + 10\%(A2\_3+C1\_0)))$.

It is important to remark that our focus is to reduce the performance gap between training and enroll-free inference scenarios. By augmenting the speaker embeddings during training, the model learns about real-world noisy inference conditions.

\vspace{-2mm}
\subsection{Overlapping Spectral Loss}
\label{sec:ov_loss}
\vspace{-2mm}
To improve the temporal coherence of the separated speech, and average out the artifacts between separated segments, we introduce the Overlapping Spectral Loss (OSL). OSL is defined as:
\vspace{-5mm}
\begin{equation}
\centering
\mathcal{L}_{overlap} = \frac{1}{K} \sum_{k=1}^{K} \sum_{f=1}^{F} \sum_{t=1}^{T} w(f,t) | \hat{Y}_k(f,t) - Y_k(f,t) |_{p}  
\end{equation}
where $\hat{Y}_k(f,t)$ is the predicted spectrogram at frequency $f$ and time $t$, $Y_k(f,t)$ is the related ground-truth spectrogram, $|\cdot |_p$ is the $p$-th order norm (e.g., $p = 1$ for L1 loss, $p = 2$ for L2 loss), $w_k(f,t)$ is a weighting function that emphasizes overlapping spectral regions, defined as:
\vspace{-5mm}
\begin{equation}
w_k(f,t) = \frac{\sum_{k=1}^{K} |Y_{k}(f,t)|}{\max_{k} |Y_{k}(f,t)| + \epsilon}
\label{eq:ov}
\end{equation}
where $k$ is the speaker index, $K$ is the total number of speakers, and $\epsilon$ prevents the denominator from being zero.
\vspace{-2mm}
\subsection{Speaker Overlap Aware Segmentation}
\vspace{-2mm}

During training, embeddings are obtained as described in Section \ref{sec:noisy_embed}. However, during evaluation, the embeddings are extracted from the centroids of speaker clusters \cite{boeddeker2024ts}. Speech utterances are extracted every 2 seconds to generate embeddings and they are used to create speaker clusters via SC \cite{ng2001spectral}. Ideally, utterances from the same speaker should be in close proximity within the same clusters. However, if speech utterances from overlapping speakers are used, the composite embeddings tend to deviate from the target speaker cluster \cite{li2020compositional}. To achieve high-quality speaker centroids representing a single speaker, the speech utterances should come from non-overlapping segments \cite{li2020compositional}. We introduced in the pipeline an overlap-aware speaker segmenter \cite{Bredin2021EndtoendSS}, which predicts whether a speech frame contains a single speaker or multiple speakers. For speaker clustering, only single speaker frames are used.


 \vspace{-3mm}

\section{Experiment}
\label{sec:experiment}
\vspace{-3mm}
\subsection{Data}
\vspace{-2mm}
\label{sec:data}
Our training dataset was created from the ``LibriSpeech Clean Train'', consisting of 460 hours of clean speech.  The mixtures were created by following a similar approach to JSALT~\cite{Raj2020IntegrationOS}, where utterances from different speakers were mixed with a maximum of 80\% overlap, to crate a total of 30k samples. Each sample contained up to 8 speakers per mixture, and was at least 60 seconds long. During training, 37-second audio chunks are taken from each sample. Different from \cite{boeddeker2024ts}, we did not use superimposition to increase the number of active speakers; instead, we decided to increase the inter-utterance overlap to 80\%. The ground truth for Stage 1 training was created with Silero-VAD \cite{SileroVAD}. For evaluation, we used LibriCSS, similar to other SOTA studies \cite{9053426}, \cite{boeddeker2024ts}. 

\vspace{-2mm}
\subsection{Model Pipeline}
\vspace{-2mm}
The base speaker-dependent VAD and speech separation model architecture follows \cite{boeddeker2024ts}, where the TS-SEP network comprises three components connected via stacking operations \cite{TSVAD} (see Figure \ref{fig:embed_aug}). The speaker embeddings are computed by using a 6 million parameter ECAPA-TDNN \cite{desplanques2020ecapatdnn} as speaker encoder. These speaker embeddings are concatenated with their corresponding representation from the speaker independent layers of TS-SEP. The second TS-SEP network component then processes each speaker individually, while sharing parameters. 
\vspace{-2mm}
\subsection{Pretraining and Training Stages}
\vspace{-2mm}
In Stage 1, a speaker-dependent VAD system is trained to determine the voice activity of each target speaker (Section \ref{sec:stage1}). As mentioned in Section \ref{sec:noisy_embed}, speaker embeddings were sampled during training. The model was trained with V1, V2, V3, V4 with equal probability. However, models with only V1/V2/V3/V4 embeddings were also trained, for comparison. 

Models trained at Stage 1 were used for Stage 2 speech separation training initialisation, where the last layer is replicated $F$ times to acquire the desired time-frequency resolution. Also in Stage 2, the speech separation model was trained with the sampling strategies described in Section \ref{sec:noisy_embed}. The overlapping spectral loss is used with the reconstruction loss with weight between $0.05$ to $0.1$, and $0.08$ produced the best performance.

\vspace{-2mm}
\subsection{Evaluation}
\label{sec:evaluation}
\vspace{-2mm}
The speech separation module performance is measured coherently to literature \cite{wang2019voicefilter, boeddeker2024ts}, by calculating the DER for the quality of diarization, Signal-to-Distortion-Ratio (SDR) for the quality of the separated speech signals, and cpWER for multi-speaker word error rate \cite{speechbrain, meeteval}. cpWER was calculated by using the ASR model NeMo toolkit~\cite{DBLP:journals/corr/abs-1909-09577}.
For the inference, let \( x(n) \) denote the discrete time domain audio signal. We frame $x(n)$ into segments of duration $N_{\text{frame}} = 30\text{ ms}$.
For each frame indexed by \( l \) (with time interval \([n_l, n_l+N_{\text{frame}}]\)), we apply a WebRTC-VAD that produces output $v_l=1$ if the frame contains speech, and $v_l=0$ otherwise.
Contiguous frames with \( v_l = 1 \) are grouped together to form speech segments. Speech segments are defined as $seg_j = [n_{\text{start},j},\, n_{\text{end},j}]$, where \( n_{\text{start},j} \) and \( n_{\text{end},j} \) are their start and end times, respectively. Two adjacent segments \( seg_j \) and \( seg_{j+1} \) that satisfy:

\vspace{-3mm}
\begin{equation}
n_{\text{start},j+1} - n_{\text{end},j} < 0.8 \text{s}
\end{equation}
\vspace{-3mm}

are merged into a single segment $seg_j' = \left[n_{\text{start},j},\, n_{\text{end},j+1}\right]$. Segments with duration less than \(0.5\) s are discarded.


To compensate for the relatively aggressive VAD settings, a small temporal padding (0.01 s) is applied to the start and end of each segment, to avoid cutting off brief speech transitions. Once all speech segments are identified, they are concatenated into a single audio stream, and passed to the speech separation model. By processing only the voiced regions, we mitigate artifacts introduced by large non-speech gaps. We reconstruct the full-length output by interleaving the separated segments with the silence that was removed. Meeteval~\cite{meeteval} toolkit is used to compute the DER metric with 0.25 s collar value and cpWER.

\vspace{-2mm}
\section{Results and Discussion}
\label{sec:results}
The DER, SDR and cpWER results of our speech separation models are shown in Table \ref{tab:sep_results}. Here, the presented models differ from each other based on their corresponding stage 1 speaker-dependent VAD models, and their sampling method used to generate embeddings during training. The DER of these stage 1 speaker-dependent VAD models is shown in Table \ref{tab:dir_results}. Table \ref{tab:comparison} presents the comparisons between the model variants of our proposed method and the SOTA baseline \cite{boeddeker2024ts}. The models' names have three fields, the first one (``sep/vad") specifies if it is a speech separation model or speaker-dependent VAD model, the second field (``e6m/ivector") shows the type of speaker embeddings used for the model training, and the third field (``V1/V2/V3/V4") mentions the type of sampling used for embeddings during training stages. Table \ref{tab:ablation} presents the ablation results. All models have been trained during Stage 1 (Section \ref{sec:stage1}) and Stage 2 (Section \ref{sec:stage2}) with the overlapping speaker data mentioned in Section \ref{sec:data}. For inference, the speaker embeddings are obtained from single speaker clusters after the Speaker Encoder-SC pipeline. For the baseline and ablation studies (Table \ref{tab:comparison}, \ref{tab:ablation}), \emph{xvectors-SC-ivectors} \cite{boeddeker2024ts}, and \emph{ECAPA-SC-ivectors} pipelines are used. We mainly compare with \cite{boeddeker2024ts} is joint diarization and separation model, while \cite{He2021TargetSpeakerVA} is mainly focused on target speaker VAD and diarization.


\begin{table}[]
\centering
\begin{adjustbox}{width=0.45\textwidth}
\begin{tabular}{ccccc}
\hline
\hline
\textbf{\begin{tabular}[c]{@{}c@{}} Separation \\ model\end{tabular}} & \textbf{\begin{tabular}[c]{@{}c@{}}Stage1\\ model\end{tabular}} & \textbf{DER} & \textbf{SDR} & \textbf{cpWER} \\
\hline
\textbf{sep\_e6m\_v1} & \textbf{vad\_e6m\_v1v2v3v4} & \textbf{4.74} & \textbf{17.44} & \textbf{12.2} \\
\textbf{sep\_e6m\_v4} & \textbf{vad\_e6m\_v1v2v3v4} & \textbf{5.13} & \textbf{15.11} & \textbf{12.19} \\
sep\_e6m\_v2 & vad\_e6m\_v2 & 5.72 & 13.51 & 13.38 \\
sep\_e6m\_v1 & vad\_e6m\_v2 & 6.06 & 15.09 & 12.25 \\
sep\_e6m\_v1 & vad\_e6m\_v1 & 6.32 & 14.23 & 14.26 \\
sep\_e6m\_v3 & vad\_e6m\_v3 & 10.98 & 10.29 & 21.87 \\
sep\_e6m\_v4 & vad\_e6m\_v4 & 5.02 & 15.23 & 12.14 \\
sep\_e6m\_v1v2v3v4 & vad\_e6m\_v2 & 5.29 & 15.89 & 12.59 \\
\textbf{sep\_e6m\_v1 + ovLoss} & \textbf{vad\_e6m\_v1v2v3v4} & \textbf{4.21} & \textbf{16.62} & \textbf{9.52} \\
\hline
\end{tabular}
\end{adjustbox}
\caption{Enrollment-free speech separation evaluation results on LibriCSS dataset with six million ECAPA embeddings + SC and Stage 2 speech separation model.}
\label{tab:sep_results}
\vspace{-5mm}
\end{table}

The results presented in Table \ref{tab:sep_results} demonstrate that the separation models achieve better performance when their training initiate from a speaker-dependent VAD, which was previously conditioned upon partial speaker-overlapping embeddings (such as those labeled as V2, V3, and V4). Stage 1 model with silence augmentation (sep\_e6m\_V3) does not performs well when V3 being used as solo embedding strategy. Moreover, we further improve 11\% the DER and 21\% cpWER results by using the OSL (Section \ref{sec:ov_loss}). OSL improves temporal coherence between the overlapping windows, thus reducing the artifacts among output speaker channels. However, some frequency bins are smoothed out causing loss of sharpness of the separated speech (hence the slight reduction of SDR). 

\emph{sep\_e6m\_v1} pretrained with \emph{vad\_e6m\_v1v2v3v4} obtained 25\%, 22.5\% and 14.4\% DER, SDR and cpWER improvement respectively over \emph{sep\_e6m\_v1} pretrained with \emph{vad\_e6m\_v1}. Combined with that, the results in Table \ref{tab:sep_results} clearly proves our initial hypothesis that pre-training the speech separation model towards \emph{speaker-overlapped noisy Target Speaker Voice Activity Detection task}, by sampling noisy (speaker overlapped) embeddings, would significantly improve speech separation and diarization tasks (Section \ref{sec:intro}, \ref{sec:noisy_embed}). Table \ref{tab:comparison} shows comparisons among the baseline SOTA model \cite{boeddeker2024ts} and our proposed models. By comparing the proposed combination of training, modelling and speaker clustering with the baseline TS-SEP pipeline \cite{boeddeker2024ts}, Table \ref{tab:comparison} shows an absolute 71.34\% DER improvement on LibriCSS, considering results with single microphone. This clearly demonstrates our significant improvements, over SOTA. Table~\ref{tab:comparison} also reports the DER achieved by the original TS-VAD~\cite{He2021TargetSpeakerVA}: it was better than TS-SEP DER, due to compromises needed to achieve both great DER and cpWER, as discussed in~\cite{boeddeker2024ts}.



\begin{table}[]
\centering
\begin{adjustbox}{width=0.25\textwidth}
\begin{tabular}{ccc}
\hline
\hline
\textbf{\begin{tabular}[c]{@{}c@{}}Stage 1\\ model\end{tabular}} & \textbf{\begin{tabular}[c]{@{}c@{}}Model\\ type\end{tabular}} & \textbf{DER} \\
\hline
vad\_e6m\_v1v2v3v4 & VAD & 10.13 \\
vad\_e6m\_v1 & VAD & 6.12 \\
vad\_e6m\_v2 & VAD & 7.26 \\
vad\_e6m\_v3 & VAD & 11.2 \\
vad\_e6m\_v4 & VAD & 8.15 \\
\hline
\end{tabular}
\end{adjustbox}
\caption{Enrollment-free speech diarization evaluation results on LibriCSS dataset with six million ECAPA embeddings + SC and Stage 1 speech diarization model.}
\label{tab:dir_results}
\vspace{-10mm}
\end{table}

The DERs of the speaker-dependent VAD models in Table \ref{tab:dir_results} shows that \emph{vad\_e6m\_v1v2v3v4} has worse DER and \emph{vad\_e6m\_v1} the best. This shows that more noisy-speaker-overlapped trained speaker-dependent VADs facilitate better representation learning in Stage 2 training, while underperforming at Stage 1. So, when the Stage 1 model learns speaker-dependent voiced/unvoiced regions with the target speaker embedding and minor overlapping speaker, it becomes adept at distinguishing primary speakers through their characteristics, while filtering out secondary ones effectively. These findings demonstrate that incorporating speaker-overlap samples for speaker embeddings in early-stage training fosters stronger adaptability and generalisation across diverse speaker mixture scenarios compared to relying solely on single-target speaker embeddings.

\begin{table}[]
\centering
\begin{adjustbox}{width=0.38\textwidth}
\begin{tabular}{cclc}
\hline
\hline
\textbf{Model} & \textbf{\begin{tabular}[c]{@{}c@{}}Model\\ type\end{tabular}} & \textbf{\begin{tabular}[c]{@{}l@{}}Evaluation\\ strategy\end{tabular}} & \textbf{DER} \\
\hline
TS-VAD BF+Masking \cite{boeddeker2024ts} & VAD & \begin{tabular}[c]{@{}l@{}}xvector + SC\\ + i-vectors\end{tabular} & 16.28 \\
\hline
TS-SEP BF+Masking \cite{boeddeker2024ts} & SEP & \begin{tabular}[c]{@{}l@{}}xvector + SC\\ + i-vectors\end{tabular} & 14.69 \\
\hline
TS-VAD \cite{He2021TargetSpeakerVA} & VAD & \begin{tabular}[c]{@{}l@{}}xvector + SC\end{tabular} & 7.5 \\
\hline
\textbf{sep\_e6m\_v1} & \textbf{SEP} & \begin{tabular}[c]{@{}l@{}}ECAPA + SC\end{tabular} & \textbf{4.74} \\
\hline
\textbf{sep\_e6m\_v1 + ov\_loss}  & \textbf{SEP} & \begin{tabular}[c]{@{}l@{}}ECAPA + SC\end{tabular} & \textbf{4.21} \\
\hline
sep\_e6m\_v4 & SEP & \begin{tabular}[c]{@{}l@{}}ECAPA + SC\end{tabular} & 5.13 \\
\hline
vad\_e6m\_v2 & VAD & \begin{tabular}[c]{@{}l@{}}ECAPA + SC \end{tabular} & 7.26 \\
\hline
\textbf{vad\_e6m\_v1} & \textbf{VAD} & \begin{tabular}[c]{@{}l@{}}ECAPA + SC\\ \end{tabular} & \textbf{6.12} \\
\hline
\end{tabular}
\end{adjustbox}
\caption{Enrollment-free speech diarization evaluation results on LibriCSS \& SOTA enrollment-free target speech separation.}
\label{tab:comparison}
\vspace{-5mm}
\end{table}


\begin{table}[]
\centering
\begin{adjustbox}{width=0.35\textwidth}
\begin{tabular}{ccccc}
\hline
\hline
\textbf{Model} & \textbf{\begin{tabular}[c]{@{}c@{}}Model\\ type\end{tabular}} & \textbf{\begin{tabular}[c]{@{}c@{}}Evaluation \\ strategy\end{tabular}} & \textbf{DER} & \textbf{cpWER} \\
\hline
sep\_ivector & SEP & oracle i-vector & 8.08 & 17.46 \\
sep\_ivector & SEP & \begin{tabular}[c]{@{}c@{}}xvector + SC \\ + ivector\end{tabular} & 17.05 & 31.26 \\
sep\_ivector & SEP & \begin{tabular}[c]{@{}c@{}}ECAPA-TDNN\\ + SC + ivector\end{tabular} & 12.16 & 24.12 \\
\hline
\end{tabular}
\end{adjustbox}
\caption{Enrollment-free speech diarization and separation results on LibriCSS, with model trained with ivectors \& our training data, and evaluated with different evaluation pipeline.}
\label{tab:ablation}
\vspace{-9mm}
\end{table}


Boeddeker \textit{et al}. found that the initial clustering for choosing single speaker utterances was better with xvector-based embeddings, while target speaker separation was better with ivector-based embeddings \cite{boeddeker2024ts}. Therefore, they employed a hybrid xvector-SC-ivector\footnote[1]{https://huggingface.co/speechbrain/spkrec-xvect-voxceleb} approach \cite{boeddeker2024ts}. However, from our study, we have found the smaller 6-million parameter ECAPA-TDNN\footnote[2]{https://huggingface.co/yangwang825/ecapa-tdnn-vox2} to outperform their hybrid approach (Table~\ref{tab:comparison}). Additionally, experiments with larger 20-million-parameter ECAPA-TDNN variants revealed marginal enhancements (+0.8--1.0\% absolute improvement) in DER. Nonetheless, due to computational efficiency, we opted for the smaller configuration.
\vspace{-3mm}
\subsection{Ablation}
\vspace{-2mm}
During the ablation study (Table \ref{tab:ablation}), we compare the xvector-ivector-based pipeline and ECAPA-TDNN-based pipeline, by training TS-SEP and TS-VAD with global i-vectors embeddings, similar to \cite{boeddeker2024ts}, by using the overlapped data described at Section \ref{sec:data}. The model is evaluated with: oracle i-vectors, \emph{ECAPA-TDNN + SC + i-vector} hybrid, and \emph{xvector + SC+ i-vectors} hybrid. The results at Table \ref{tab:ablation} show that we could achieve similar numbers to what they reported in their original paper (\emph{xvector + SC+ i-vectors})~\cite{boeddeker2024ts}. Furthermore, we show at Table \ref{tab:ablation}, row 3 that ECAPA-TDNN embedding based approach is more accurate (improving 28.6\% on DER and 22\% on cpWER) than the xvector-ivector hybrid in the presence of overlapping speakers, while using the same \emph{sep\_ivector} model.

\vspace{-3mm}
\section{Conclusions}
\vspace{-2mm}
In this paper, we proposed a novel embedding sampling method for training, and overlapping spectral loss for enrollment-free target speech separation. The proposed model outperformed the previous SOTA by 71\% in DER and 69\% in cpWER (the 69\% relative cpWER improvement is calculated  by comparing Table 4 row 2 with Table 1 row 9). Future work will control the embedding sample overlaps for more robust representation learning and improve overlap focused selective window spectral loss, for further improvement on the overlapped speech regions.
\label{sec:conclusion}

\vspace{-7mm}
\bibliographystyle{IEEEtran}

\bibliography{strings}

\end{document}